\pdfoutput=1 
\documentclass[sigconf,screen]{acmart}

\renewcommand\footnotetextcopyrightpermission[1]{} 
\setcopyright{none}
\acmConference[arXiv preprint]{}{}{}
\thanks{This version is a preprint submitted to arXiv.}

\usepackage{graphicx}
\usepackage{booktabs}
\usepackage{enumitem}
\usepackage{verbatim}

\usepackage{tikz}
\usepackage{pgfplots}
\pgfplotsset{compat=1.18}
\usetikzlibrary{patterns, positioning, calc}
\pgfdeclarelayer{background}
\pgfdeclarelayer{foreground}
\pgfsetlayers{background,main,foreground}
\usepackage{listings}
\definecolor{kit-green}{RGB}{0, 150, 130}
\definecolor{kit-green100}{RGB}{0, 150, 130}
\definecolor{kit-green90}{rgb}{0.1, 0.6294, 0.5588}
\definecolor{kit-green80}{rgb}{0.2, 0.6706, 0.6078}
\definecolor{kit-green75}{rgb}{0.25, 0.6912, 0.6324}
\definecolor{kit-green70}{rgb}{0.3, 0.7118, 0.6569}
\definecolor{kit-green60}{rgb}{0.4, 0.7529, 0.7059}
\definecolor{kit-green50}{rgb}{0.5, 0.7941, 0.7549}
\definecolor{kit-green40}{rgb}{0.6, 0.8353, 0.8039}
\definecolor{kit-green30}{rgb}{0.7, 0.8765, 0.8529}
\definecolor{kit-green25}{rgb}{0.75, 0.8971, 0.8775}
\definecolor{kit-green20}{rgb}{0.8, 0.9176, 0.902}
\definecolor{kit-green15}{rgb}{0.85, 0.9382, 0.9265}
\definecolor{kit-green10}{rgb}{0.9, 0.9588, 0.951}
\definecolor{kit-green5}{rgb}{0.95, 0.9794, 0.9755}

\definecolor{kit-blue}{RGB}{70, 100, 170}
\definecolor{kit-blue100}{RGB}{70, 100, 170}
\definecolor{kit-blue90}{rgb}{0.3471, 0.4529, 0.7}
\definecolor{kit-blue80}{rgb}{0.4196, 0.5137, 0.7333}
\definecolor{kit-blue75}{rgb}{0.4559, 0.5441, 0.75}
\definecolor{kit-blue70}{rgb}{0.4922, 0.5745, 0.7667}
\definecolor{kit-blue60}{rgb}{0.5647, 0.6353, 0.8}
\definecolor{kit-blue50}{rgb}{0.6373, 0.6961, 0.8333}
\definecolor{kit-blue40}{rgb}{0.7098, 0.7569, 0.8667}
\definecolor{kit-blue30}{rgb}{0.7824, 0.8176, 0.9}
\definecolor{kit-blue25}{rgb}{0.8186, 0.848, 0.9167}
\definecolor{kit-blue20}{rgb}{0.8549, 0.8784, 0.9333}
\definecolor{kit-blue15}{rgb}{0.8912, 0.9088, 0.95}
\definecolor{kit-blue10}{rgb}{0.9275, 0.9392, 0.9667}
\definecolor{kit-blue5}{rgb}{0.9637, 0.9696, 0.9833}

\definecolor{kit-red}{RGB}{162, 34, 35}
\definecolor{kit-red100}{RGB}{162, 34, 35}
\definecolor{kit-red90}{rgb}{0.6718, 0.22, 0.2235}
\definecolor{kit-red80}{rgb}{0.7082, 0.3067, 0.3098}
\definecolor{kit-red75}{rgb}{0.7265, 0.35, 0.3529}
\definecolor{kit-red70}{rgb}{0.7447, 0.3933, 0.3961}
\definecolor{kit-red60}{rgb}{0.7812, 0.48, 0.4824}
\definecolor{kit-red50}{rgb}{0.8176, 0.5667, 0.5686}
\definecolor{kit-red40}{rgb}{0.8541, 0.6533, 0.6549}
\definecolor{kit-red30}{rgb}{0.8906, 0.74, 0.7412}
\definecolor{kit-red25}{rgb}{0.9088, 0.7833, 0.7843}
\definecolor{kit-red20}{rgb}{0.9271, 0.8267, 0.8275}
\definecolor{kit-red15}{rgb}{0.9453, 0.87, 0.8706}
\definecolor{kit-red10}{rgb}{0.9635, 0.9133, 0.9137}
\definecolor{kit-red5}{rgb}{0.9818, 0.9567, 0.9569}

\definecolor{kit-yellow}{RGB}{252, 229, 0}
\definecolor{kit-yellow100}{RGB}{252, 229, 0}
\definecolor{kit-yellow90}{rgb}{0.9894, 0.9082, 0.1}
\definecolor{kit-yellow80}{rgb}{0.9906, 0.9184, 0.2}
\definecolor{kit-yellow75}{rgb}{0.9912, 0.9235, 0.25}
\definecolor{kit-yellow70}{rgb}{0.9918, 0.9286, 0.3}
\definecolor{kit-yellow60}{rgb}{0.9929, 0.9388, 0.4}
\definecolor{kit-yellow50}{rgb}{0.9941, 0.949, 0.5}
\definecolor{kit-yellow40}{rgb}{0.9953, 0.9592, 0.6}
\definecolor{kit-yellow30}{rgb}{0.9965, 0.9694, 0.7}
\definecolor{kit-yellow25}{rgb}{0.9971, 0.9745, 0.75}
\definecolor{kit-yellow20}{rgb}{0.9976, 0.9796, 0.8}
\definecolor{kit-yellow15}{rgb}{0.9982, 0.9847, 0.85}
\definecolor{kit-yellow10}{rgb}{0.9988, 0.9898, 0.9}
\definecolor{kit-yellow5}{rgb}{0.9994, 0.9949, 0.95}

\definecolor{kit-orange}{RGB}{223, 155, 27}
\definecolor{kit-orange100}{RGB}{223, 155, 27}
\definecolor{kit-orange90}{rgb}{0.8871, 0.6471, 0.1953}
\definecolor{kit-orange80}{rgb}{0.8996, 0.6863, 0.2847}
\definecolor{kit-orange75}{rgb}{0.9059, 0.7059, 0.3294}
\definecolor{kit-orange70}{rgb}{0.9122, 0.7255, 0.3741}
\definecolor{kit-orange60}{rgb}{0.9247, 0.7647, 0.4635}
\definecolor{kit-orange50}{rgb}{0.9373, 0.8039, 0.5529}
\definecolor{kit-orange40}{rgb}{0.9498, 0.8431, 0.6424}
\definecolor{kit-orange30}{rgb}{0.9624, 0.8824, 0.7318}
\definecolor{kit-orange25}{rgb}{0.9686, 0.902, 0.7765}
\definecolor{kit-orange20}{rgb}{0.9749, 0.9216, 0.8212}
\definecolor{kit-orange15}{rgb}{0.9812, 0.9412, 0.8659}
\definecolor{kit-orange10}{rgb}{0.9875, 0.9608, 0.9106}
\definecolor{kit-orange5}{rgb}{0.9937, 0.9804, 0.9553}

\definecolor{kit-lightgreen}{RGB}{140, 182, 60}
\definecolor{kit-lightgreen100}{RGB}{140, 182, 60}
\definecolor{kit-lightgreen90}{rgb}{0.5941, 0.7424, 0.3118}
\definecolor{kit-lightgreen80}{rgb}{0.6392, 0.771, 0.3882}
\definecolor{kit-lightgreen75}{rgb}{0.6618, 0.7853, 0.4265}
\definecolor{kit-lightgreen70}{rgb}{0.6843, 0.7996, 0.4647}
\definecolor{kit-lightgreen60}{rgb}{0.7294, 0.8282, 0.5412}
\definecolor{kit-lightgreen50}{rgb}{0.7745, 0.8569, 0.6176}
\definecolor{kit-lightgreen40}{rgb}{0.8196, 0.8855, 0.6941}
\definecolor{kit-lightgreen30}{rgb}{0.8647, 0.9141, 0.7706}
\definecolor{kit-lightgreen25}{rgb}{0.8873, 0.9284, 0.8088}
\definecolor{kit-lightgreen20}{rgb}{0.9098, 0.9427, 0.8471}
\definecolor{kit-lightgreen15}{rgb}{0.9324, 0.9571, 0.8853}
\definecolor{kit-lightgreen10}{rgb}{0.9549, 0.9714, 0.9235}
\definecolor{kit-lightgreen5}{rgb}{0.9775, 0.9857, 0.9618}

\definecolor{kit-purple}{RGB}{163, 16, 124}
\definecolor{kit-purple100}{RGB}{163, 16, 124}
\definecolor{kit-purple90}{rgb}{0.6753, 0.1565, 0.5376}
\definecolor{kit-purple80}{rgb}{0.7114, 0.2502, 0.589}
\definecolor{kit-purple75}{rgb}{0.7294, 0.2971, 0.6147}
\definecolor{kit-purple70}{rgb}{0.7475, 0.3439, 0.6404}
\definecolor{kit-purple60}{rgb}{0.7835, 0.4376, 0.6918}
\definecolor{kit-purple50}{rgb}{0.8196, 0.5314, 0.7431}
\definecolor{kit-purple40}{rgb}{0.8557, 0.6251, 0.7945}
\definecolor{kit-purple30}{rgb}{0.8918, 0.7188, 0.8459}
\definecolor{kit-purple25}{rgb}{0.9098, 0.7657, 0.8716}
\definecolor{kit-purple20}{rgb}{0.9278, 0.8125, 0.8973}
\definecolor{kit-purple15}{rgb}{0.9459, 0.8594, 0.9229}
\definecolor{kit-purple10}{rgb}{0.9639, 0.9063, 0.9486}
\definecolor{kit-purple5}{rgb}{0.982, 0.9531, 0.9743}

\definecolor{kit-brown}{RGB}{167, 130, 46}
\definecolor{kit-brown100}{RGB}{167, 130, 46}
\definecolor{kit-brown90}{rgb}{0.6894, 0.5588, 0.2624}
\definecolor{kit-brown80}{rgb}{0.7239, 0.6078, 0.3443}
\definecolor{kit-brown75}{rgb}{0.7412, 0.6324, 0.3853}
\definecolor{kit-brown70}{rgb}{0.7584, 0.6569, 0.4263}
\definecolor{kit-brown60}{rgb}{0.7929, 0.7059, 0.5082}
\definecolor{kit-brown50}{rgb}{0.8275, 0.7549, 0.5902}
\definecolor{kit-brown40}{rgb}{0.862, 0.8039, 0.6722}
\definecolor{kit-brown30}{rgb}{0.8965, 0.8529, 0.7541}
\definecolor{kit-brown25}{rgb}{0.9137, 0.8775, 0.7951}
\definecolor{kit-brown20}{rgb}{0.931, 0.902, 0.8361}
\definecolor{kit-brown15}{rgb}{0.9482, 0.9265, 0.8771}
\definecolor{kit-brown10}{rgb}{0.9655, 0.951, 0.918}
\definecolor{kit-brown5}{rgb}{0.9827, 0.9755, 0.959}

\definecolor{kit-cyan}{RGB}{35, 161, 224}
\definecolor{kit-cyan100}{RGB}{35, 161, 224}
\definecolor{kit-cyan90}{rgb}{0.2235, 0.6682, 0.8906}
\definecolor{kit-cyan80}{rgb}{0.3098, 0.7051, 0.9027}
\definecolor{kit-cyan75}{rgb}{0.3529, 0.7235, 0.9088}
\definecolor{kit-cyan70}{rgb}{0.3961, 0.742, 0.9149}
\definecolor{kit-cyan60}{rgb}{0.4824, 0.7788, 0.9271}
\definecolor{kit-cyan50}{rgb}{0.5686, 0.8157, 0.9392}
\definecolor{kit-cyan40}{rgb}{0.6549, 0.8525, 0.9514}
\definecolor{kit-cyan30}{rgb}{0.7412, 0.8894, 0.9635}
\definecolor{kit-cyan25}{rgb}{0.7843, 0.9078, 0.9696}
\definecolor{kit-cyan20}{rgb}{0.8275, 0.9263, 0.9757}
\definecolor{kit-cyan15}{rgb}{0.8706, 0.9447, 0.9818}
\definecolor{kit-cyan10}{rgb}{0.9137, 0.9631, 0.9878}
\definecolor{kit-cyan5}{rgb}{0.9569, 0.9816, 0.9939}

\definecolor{kit-gray}{RGB}{0, 0, 0}
\definecolor{kit-gray100}{RGB}{0, 0, 0}
\definecolor{kit-gray90}{rgb}{0.1, 0.1, 0.1}
\definecolor{kit-gray80}{rgb}{0.2, 0.2, 0.2}
\definecolor{kit-gray75}{rgb}{0.25, 0.25, 0.25}
\definecolor{kit-gray70}{rgb}{0.3, 0.3, 0.3}
\definecolor{kit-gray60}{rgb}{0.4, 0.4, 0.4}
\definecolor{kit-gray50}{rgb}{0.5, 0.5, 0.5}
\definecolor{kit-gray40}{rgb}{0.6, 0.6, 0.6}
\definecolor{kit-gray30}{rgb}{0.7, 0.7, 0.7}
\definecolor{kit-gray25}{rgb}{0.75, 0.75, 0.75}
\definecolor{kit-gray20}{rgb}{0.8, 0.8, 0.8}
\definecolor{kit-gray15}{rgb}{0.85, 0.85, 0.85}
\definecolor{kit-gray10}{rgb}{0.9, 0.9, 0.9}
\definecolor{kit-gray5}{rgb}{0.95, 0.95, 0.95}

\usepackage[theorems, skins, breakable, listings]{tcolorbox}
\usepackage{footnote}

\newtcbtheorem[use counter=prompt]{promptL}{Prompt}{%
	attach title to upper,%
	after title={\ },%
	colback=white,%
	colframe=kit-blue50,%
	fonttitle={\bfseries\color{black}},%
        sharp corners=east,%
        leftrule=3mm,
	subtitle style={%
		boxrule=0.4pt,%
		colback=kit-blue70,%
		fonttitle={\bfseries\color{white}}%
	}%
}{prompt}

\BeforeBeginEnvironment{promptL}{\savenotes}
\AfterEndEnvironment{promptL}{\spewnotes}

\newtcbtheorem[use counter=prompt]{promptR}{Prompt}{%
	attach title to upper,%
	after title={\ },%
	colback=white,%
	colframe=kit-blue50,%
	fonttitle={\bfseries\color{black}},%
        sharp corners=west,%
        rightrule=3mm,%
	subtitle style={%
		boxrule=0.4pt,%
		colback=kit-blue70,%
		fonttitle={\bfseries\color{white}}%
	}%
}{prompt}
\BeforeBeginEnvironment{promptR}{\savenotes}
\AfterEndEnvironment{promptR}{\spewnotes}

\AtBeginDocument{%
  }

\setcopyright{acmlicensed}
\copyrightyear{2018}
\acmYear{2018}
\acmDOI{XXXXXXX.XXXXXXX}

\acmISBN{978-1-4503-XXXX-X/2018/06}

\begin{document}




\title{An Experience Report on a Pedagogically Controlled, Curriculum-Constrained AI Tutor for SE Education}

\author{Lucia Happe}
\orcid{0000-0003-1117-6880}
\email{lucia.happe@kit.edu}
\affiliation{%
  \institution{Karlsruhe Institute of Technology}
  \city{Karlsruhe}
  \country{Germany}
}

\author{Dominik Fuchß}
\orcid{0000-0001-6410-6769}
\email{dominik.fuchss@kit.edu}
\affiliation{%
  \institution{Karlsruhe Institute of Technology}
  \city{Karlsruhe}
  \country{Germany}
}

\author{Luca Hüttner}
\orcid{XXXX-XXXX-XXXX-XXXX}
\email{luca.huettner@student.kit.edu}
\affiliation{%
  \institution{Karlsruhe Institute of Technology}
  \city{Karlsruhe}
  \country{Germany}
}

\author{Kai Marquardt}
\orcid{0000-0003-0422-5314}
\email{kai.marquardt@alumni.kit.edu}
\affiliation{%
  \institution{Karlsruhe Institute of Technology}
  \city{Karlsruhe}
  \country{Germany}
}

\author{Anne Koziolek}
\orcid{0000-0002-1593-3394}
\email{koziolek@kit.edu}
\affiliation{%
  \institution{Karlsruhe Institute of Technology}
  \city{Karlsruhe}
  \country{Germany}
}


\renewcommand{\shortauthors}{Happe et al.}

\begin{abstract}
The integration of artificial intelligence (AI) into education continues to evoke both promise and skepticism. While past waves of technological optimism often fell short, recent advances in large language models (LLMs) have revived the vision of scalable, individualized tutoring. This paper presents the design and pilot evaluation of \textit{RockStartIT Tutor}, an AI-powered assistant developed for a digital programming and computational thinking course within the RockStartIT initiative. Powered by GPT-4 via OpenAI’s Assistant API, the tutor employs a novel prompting strategy and a modular, semantically tagged knowledge base to deliver context-aware, personalized, and curriculum-constrained support for secondary school students.

We evaluated the system using the Technology Acceptance Model (TAM) with 13 students and teachers. Learners appreciated the low-stakes environment for asking questions and receiving scaffolded guidance. Educators emphasized the system’s potential to reduce cognitive load during independent tasks and complement classroom teaching. Key challenges include prototype limitations, a small sample size, and the need for long-term studies with the target age group.

Our findings highlight a pragmatic approach to AI integration that requires no model training—using structure and prompts to shape behavior. We position AI tutors not as teacher replacements but as enabling tools that extend feedback access, foster inquiry, and support what schools do best: help students learn.

\end{abstract}

\ccsdesc[500]{Applied computing~Education}
\ccsdesc[500]{Social and professional topics~Computer science education}
\ccsdesc[500]{Social and professional topics~Computer engineering education}
\ccsdesc[500]{Social and professional topics~Software engineering education}
\ccsdesc[500]{Social and professional topics~Computing education}

\keywords{AI tutors, GPT-4, chatbots, LLMs, software engineering education, technology acceptance model, digital learning}



\maketitle
\section{Introduction}
\label{sec:Introduction}

In 1913, Thomas Edison famously predicted that motion pictures would revolutionize education and render books in schools obsolete~\cite{saettler1990history}. A century later, similar optimism surrounds artificial intelligence (AI) in education. Modern AI-powered chatbots have revived long-standing visions of a personal tutor for every student—capable of answering questions, guiding inquiry, and personalizing instruction at scale. Yet, as with past technological interventions, the transformative promise of AI often falters in practice—due to limited evaluation, poor classroom integration, or misalignment with pedagogical goals~\cite{cuban2001oversold, reich2020failure}.

Despite these challenges, AI retains significant potential, especially when applied thoughtfully to support rather than replace instruction. In contexts where teacher attention is limited and learning is increasingly digital and asynchronous, AI tutors can fill critical gaps. This is particularly relevant in software engineering (SE) education, where students frequently work independently and face complex problem-solving tasks without immediate support~\cite{gross2022aiSEsupport}. Initiatives like RockStartIT~\cite{happe2023interdisciplinary} show that engagement in SE increases when computing is taught through real-world, interdisciplinary problems. However, maintaining that engagement requires more than content delivery—it demands responsive guidance.

AI tutors offer scalable, individualized support that is always available, nonjudgmental, and capable of providing immediate feedback~\cite{winkler2020chatboted}. Such systems may lower psychological barriers and encourage low-stakes, exploratory learning~\cite{holstein2019studentai}. Especially in SE education—where self-study is the norm—AI tutors can be a valuable alternative to learning in isolation.

Still, as Reich notes, “talking to robots is boring”~\cite{reich2020failure}. Learning thrives in social, human-centered environments. AI tutors cannot replace collaborative classroom discourse, peer interaction, or teacher mentorship. However, in self-paced SE environments, they can offer a meaningful improvement over silence. Moreover, dialogue with AI can foster metacognitive skills—such as articulating reasoning, formulating questions, and iteratively refining ideas—that are essential to SE practice~\cite{zhu2021aiSEskills}.

Rather than proposing AI tutors as replacements for educators, this work embraces a complementary vision: AI as an enabler that supports learner autonomy, amplifies instructional reach, and allows teachers to focus on what they do best—facilitating insight, feedback, and human connection~\cite{luckin2016intelligenceunleashed}.

This paper presents the design and evaluation of a pedagogically controlled AI tutor developed for a RockStartIT course on programming and computational thinking. The tutor uses OpenAI’s GPT-4 model via the Assistant API, embedded in a device-independent web interface. Instead of relying on model fine-tuning or external search infrastructure, the system combines a modular, semantically tagged knowledge base with carefully engineered prompts to control behavior, guide dialogue, and restrict output to approved instructional content. This approach enables safe, personalized, and curriculum-aligned tutoring without requiring any custom training or internal model modifications.

Our goal is not to replace educators but to offer a lightweight and pragmatic architecture that empowers teachers and learners alike. The AI tutor serves as an enabler: filling gaps in individual support, reducing learner hesitation, and promoting inquiry-driven learning—particularly in asynchronous and independent study contexts common in SE education. We evaluate the system through the lens of the Technology Acceptance Model (TAM)~\cite{davis_perceived_1989}, combining quantitative and qualitative feedback from both students and teachers. Our findings inform short-term design improvements and point toward future work in adaptive scaffolding, trust calibration, and broader curricular integration.

\section{Background and Related Work}
\label{sec:background}

In this section, we provide an overview of the relevant background and related work that informs our research on AI tutors in secondary software engineering education.

\subsection{AI Assistants for Education}
Large language models (LLMs) such as GPT-3.5 and GPT-4 have gained traction as tools for supporting programming and computational thinking education. These AI-powered assistants enable natural language interaction and can function as virtual tutors, providing timely, context-sensitive feedback across a wide range of tasks.

\citet{dai_2023} explored the use of ChatGPT to provide automated feedback to students. Their study evaluated the readability, instructional similarity, and learning effectiveness of ChatGPT's responses. The results showed that ChatGPT generated coherent and clear feedback aligned with human instructor evaluations and offered insights into student thinking, potentially improving learning outcomes.

Similarly,~\citet{frankford_ai-tutoring_2024} developed an AI tutor based on GPT-3.5-Turbo to support students in software engineering exercises, aiming to reduce the workload on teaching assistants. Integrated into the APAS Artemis system, the AI tutor offered benefits such as faster response times and 24/7 availability. However, students' perceptions of helpfulness varied, and concerns were raised regarding generic responses and their potential impact on deep learning.

\citet{bassner_2024} introduced Iris, an AI tutor incorporating chain-of-thought prompting and access to student code and tasks. User studies showed that Iris clarified concepts and provided relevant hints without revealing answers, reinforcing its complementary role to human instructors.

\citet{hellas_2023} examined the use of GPT-3.5 and Codex for code feedback. While the models detected at least one genuine issue in 90\% of cases, they also produced false positives and often failed to identify all problems. Instructors frequently needed to refine AI feedback before it could be useful to students, highlighting the limitations of LLMs in their current state.

In a broader educational context,~\citet{kerstin_2025} conducted a large-scale randomized controlled trial comparing an AI tutor with active learning in physics education. The AI-supported group showed significantly greater learning gains and reported higher engagement, emphasizing the potential of well-designed AI tutors to complement traditional instruction.

While promising, most prior studies have focused on higher education and technical domains. There remains a notable gap in research on AI tutors for younger learners (aged 13--16), especially in real-world deployments where cognitive development, trust, and instructional needs differ significantly.

\subsection{Pedagogical Alignment and Instructional Control}
Despite rapid progress, many current AI tutoring systems rely on generic prompts and unstructured content, lacking alignment with educational goals. Few systems offer mechanisms to control instructional behavior, adapt scaffolding levels, or ensure that the AI adheres to curriculum boundaries. Our work addresses these limitations by embedding domain-specific instructions and using a modular, semantically tagged knowledge base to guide AI behavior.

To our knowledge, no previous study has deployed an AI tutor that uses structured tagging and traversal logic to dynamically adapt behavior while ensuring fidelity to pre-approved instructional content. This allows for fine-grained pedagogical control without requiring custom model training or retraining.

\subsection{Human-AI Collaboration and Trust Calibration}
Recent research has begun to explore how AI tutors can be integrated into classroom dynamics without displacing human educators.~\citet{holstein2019studentai} emphasize the importance of co-designing AI systems that empower teachers and support their instructional strategies. Similarly, trust calibration has emerged as a critical concern in AI-supported learning~\cite{nazaretsky2022instrument}. Learners may over-trust or under-trust AI responses, especially when system transparency is low.

Our implementation supports teacher-aligned control through behavior tagging, content restriction, and transparent response sourcing. We also collect open-ended feedback on students’ and teachers’ perceptions of trustworthiness and limitations.

\subsection{RockStartIT Interdisciplinary SE Courses}
RockStartIT is an initiative that provides engaging and accessible online SE courses for school-aged learners. Based on the IDEA (Interdisciplinary, Diverse, Exploratory, Active) framework~\cite{happe2021effective, happe2022frustrations}, these courses combine real-world challenges with computing instruction.

RockStartIT expeditions immerse students in interdisciplinary journeys that encourage inquiry, teamwork, and experimentation. For example, in the ``Save the Bees'' expedition, students investigate colony collapse disorder through web development, data analysis, and AI. They learn programming and problem-solving by working on tasks such as creating awareness websites, analyzing bee flight data, and evaluating hive health using SQL and image data. These explorations are designed to connect CS skills with meaningful contexts.

\subsection{Technology Acceptance in Educational Contexts}
The Technology Acceptance Model~(TAM)~\cite{davis_perceived_1989} is a foundational framework for understanding how users adopt educational technologies. It identifies Perceived Usefulness (PU), Perceived Ease of Use (PEOU), and Intention to Use (IU) as central factors influencing acceptance.

While TAM has been widely applied to chatbots and educational tools, most studies rely on simulated interactions or prototype systems. Few evaluate fully operational systems in authentic classroom contexts. This gap underscores the need for empirical studies that capture the real-world challenges and contextual influences on AI acceptance.

\subsection{Positioning This Work}
Our research advances the field in several ways:
\begin{itemize}
    \item We deploy a fully functional AI tutor using a modular and behavior-tagged knowledge base --- without retraining or external search.
    \item We evaluate the system in a real-world SE course with secondary students and teachers.
    \item We extend TAM-based evaluation to include open-ended perceptions of trust, adaptability, and instructional fit.
\end{itemize}

By focusing on grounded deployment and educational control, our work contributes practical insights into how AI tutors can be safely and effectively integrated into school settings.

\section{System Architecture and Implementation}
\label{sec:Implementation}

Rather than building a new AI model, the work presents a novel, lightweight, and powerful framework for turning general-purpose LLMs into pedagogically effective tutors using only clever architecture design, tool restriction, and instructional prompting.
This makes it highly replicable and safe for educational use.
The AI tutor was designed to provide pedagogically aligned, adaptive, and context-sensitive support for learners in a modular and scalable manner.
This section outlines the system architecture, key components, and the core innovations behind its implementation.

\subsection{Design Goals and Requirements}
The AI tutor was designed to support students aged 14--16 participating in the RockStartIT course. The key design requirements were as follows:
\begin{itemize}
    \item Provide context-aware, personalized support that is \textbf{strictly grounded in the course content}, and with mechanisms to \textbf{prevent breakout into unrelated topics or general knowledge}.
    \item Communicate in an age-appropriate, motivating, and accessible language suitable for lower secondary school students.
    \item Promote \textbf{inquiry and independent thinking by offering hints and guidance instead of direct solutions}.
    \item Ensure seamless access across devices without requiring student login or registration.
    \item Enable real-time interaction while \textbf{avoiding hallucinations and enforcing curriculum boundaries}, ensuring pedagogical safety and content fidelity.
\end{itemize}

\subsection{Overall Architecture}
The AI Tutor was implemented as a modular, web-based educational assistant powered by OpenAI’s GPT-4o via the Assistant API.
%
It follows a client-server model with three main components: 

\begin{itemize}
    \item \textbf{Frontend:} A device-independent chat interface built with React and Next.js, using the Chat UI Kit React library. It allows students to intuitively submit questions and displays real-time streamed responses from the AI.
    \item \textbf{Backend:} Implemented using Next.js API routes, the backend securely handles API keys, manages user sessions and conversation threads, and routes messages between the frontend and the AI model.
    \item \textbf{AI Layer:} Utilizes OpenAI’s GPT-4o via the Assistant API. It is configured with pedagogical instructions and connected to a structured knowledgebase composed of Markdown-formatted course content uploaded as file tools.
\end{itemize}

\subsection{Structured Knowledgebase Integration}

\begin{figure}[h]
\centering
\resizebox{\linewidth}{!}{
\begin{tikzpicture}[font=\scriptsize,
  node distance=1em and 2em,
  every node/.style={align=left, anchor=west, text width=18em}]

\node[font=\small] (title) {\textbf{Course Module (Markdown)}};

\node[below of=title] (tasksLabel) {\texttt{\#\# Tasks}};
\node[below of=tasksLabel] (tasksText) {"Write a function that checks if a number is even."};

\node[below of=tasksText] (hintsLabel) {\texttt{\#\# Hints}};
\node[below of=hintsLabel] (hintsText) {"Remember the modulo operator \texttt{\%}."};

\node[below of=hintsText] (expLabel) {\texttt{\#\# Explanations}};
\node[below of=expLabel] (expText) {"Even numbers have no remainder when ..."};

\node[below of=expText,yshift=-1em] (solLabel) {\texttt{\#\# Solutions (do not show)}};
\node[below of=solLabel] (solText) {\textit{"def is\_even(n): return n \% 2 == 0"}};

\node[below of=solText] (motLabel) {\texttt{\#\# Motivation}};
\node[below of=motLabel] (motText) {"You're almost there! Great thinking!"};

\node[below of=motText] (misLabel) {\texttt{\#\# Misconceptions}};
\node[below of=misLabel] (misText) {"Dividing by 2 and checking if the result ..."};

\node[right=of title,xshift=-5em,yshift=-0.75em, anchor=north west,font=\small, text width=11em] (behaviorTitle) {\textbf{AI Tutor Behavior}};

\node[below of=behaviorTitle, text width=11em] (alwaysLabel) {\textbf{Always:}};
\node[below of=alwaysLabel, text width=11em] (always1) {-- Use only course content};
\node[below of=always1, text width=11em] (always2) {-- Speak English, age-appropriate};
\node[below of=always2, text width=11em] (always3) {-- Scaffold with hints \& explanations};

\node[right=of solLabel,xshift=-5em, text width=11em] (neverLabel) {\textbf{Never:}};
\node[below of=neverLabel, text width=11em] (never1) {-- Go off-topic};
\node[below of=never1, text width=11em] (never2) {-- Reveal full solutions};
\node[below of=never2, text width=11em] (never3) {-- Answer outside the curriculum};

\draw[thick, kit-green, rounded corners]
  ($(title.north west)+(0.05,-0.45)$) rectangle ($(neverLabel.north east)+(-0.2,0.1)$);

\draw[thick, kit-red, rounded corners]
  ($(solLabel.north west)+(0.05,0)$) rectangle ($(never3.south east)+(-0.2,-0.8)$);

\begin{pgfonlayer}{background}
\draw[thick,draw=none,fill=kit-blue10,rounded corners] ($(title.north west)+(-0.5em,0)$) rectangle ($(misText.south east)+(-3em,-1em)$);
\end{pgfonlayer}
\end{tikzpicture}
}
\caption{Simplified example of tagged Markdown structure for a learning content.}
\label{fig:tagged-markdown}
\end{figure}

A core innovation of the KI Tutor lies in its use of OpenAI’s Assistant API tool-calling mechanism to strictly constrain responses to a structured, curriculum-aligned knowledgebase. This eliminates hallucinations and ensures pedagogical safety.
The knowledgebase is composed of modular, Markdown-formatted files that encapsulate the RockStart IT course content (as illustrated in Fig. \ref{fig:tagged-markdown}), including concepts, scaffolded hints, learning tasks, and motivational prompts. These files are uploaded as a dedicated \texttt{file\_search} tool accessible to the Assistant API, ensuring that all responses remain grounded in predefined instructional material.

\paragraph{Tagged Modular Format:}
The course content was organized into multiple Markdown files, each corresponding to a specific learning module. Within each file, content was segmented into pedagogically meaningful units using consistent second-level Markdown headers such as \texttt{tasks}, \texttt{hints}, \texttt{explanations}, \texttt{solutions} (explicitly marked as do not show), \texttt{motivation}, and \texttt{misconceptions}.
Please be aware that in our original study all terms were in German.

This structure served several key purposes: (i) it enabled modular, curriculum-aligned retrieval through consistent file and section formatting; (ii) it ensured instructional control by mapping each content type to defined tutor behaviors; and (iii) it supported scalable expansion of the knowledge base without requiring changes to the underlying system logic. The assistant used a layered traversal strategy: beginning with hints and explanations, escalating only when needed, and withholding direct solutions unless explicitly permitted. Motivational feedback was delivered upon completion of conceptual milestones. This scaffolding approach reflects principles from intelligent tutoring systems and supports adaptive learning paths. Unlike general RAG or fine-tuning pipelines, this implementation offers tight control over knowledge access and response generation—ensuring factual precision, pedagogical alignment, and safety without training a custom model.

\subsection{Instruction-Based Pedagogical Behavior}

To align the AI assistant’s behavior with educational goals, the system employs a carefully crafted instruction set using OpenAI's system prompt mechanism. These instructions define the assistant’s role, restrict its knowledge scope, and ensure pedagogical integrity.

\begin{itemize}
    \item Always respond in German using simple, age-appropriate language.
    \item Limit all responses strictly to the provided course content.
    \item Avoid giving direct solutions, especially for tagged answers (``nicht herausgeben!'').
    \item Support learners using Socratic questioning, scaffolded hints, and encouragement.
    \item Maintain a motivating and friendly tone appropriate for students aged 14--16.
\end{itemize}

\paragraph{Prompt Engineering and Behavior Control:}
Challenges such as hallucination, verbosity, and over-answering were addressed through iterative prompt tuning. The final system prompt (\autoref{prompt:system}) includes explicit guidance.


To align AI responses with pedagogical intent, the system incorporates semantic behavior tags (e.g., \texttt{task}, \texttt{explanation}, \texttt{solution}) within the knowledge base. These tags inform the assistant how to respond based on the instructional function of the content. When a student submits a query, the system:

\begin{enumerate}
    \item Classifies the user's intent (e.g., clarification, task assistance, conceptual exploration).
    \item Filters and retrieves content segments with matching tags.
    \item Constructs a response using layered instructional strategies tailored to the content type and learner context.
\end{enumerate}

For example, task-related queries trigger a \textit{hint-first} scaffolding strategy, while conceptual questions follow an \textit{explanation-then-analogy} pattern to deepen understanding. This combination of content-based restrictions and tag-driven instructional behavior—achieved entirely through prompt engineering—offers a lightweight yet powerful method for developing safe, adaptive, and curriculum-aligned AI tutors without requiring model fine-tuning.

\begin{promptR}{System Prompt of AI Tutor}{system}\\
This GPT is a chatbot tutor named AI Tutor, designed to assist students with questions related to an online course called "Rettet die Bienen!".
\\\\
It should provide clear, concise answers, offer additional explanations when necessary, and be patient and encouraging. It should ALWAYS and only use the Information in the Knowledgebase to answer the questions. Only answer questions relevant to the course.
\\\\
All responses should be written in German and kept concise. The students are around the age of 14 to 16 years old, so answer in that niveau.
\\\\
When asked for solutions to a question that is in the 'Antworten' part of the files in the knowledge base, AI Tutor should NEVER give out the solution directly but instead guide and motivate students to find the answers themselves by giving tips, asking guiding questions, and encouraging critical thinking without directly providing the answers.
\end{promptR}

\subsection{Context Preservation and Adaptive Interaction}

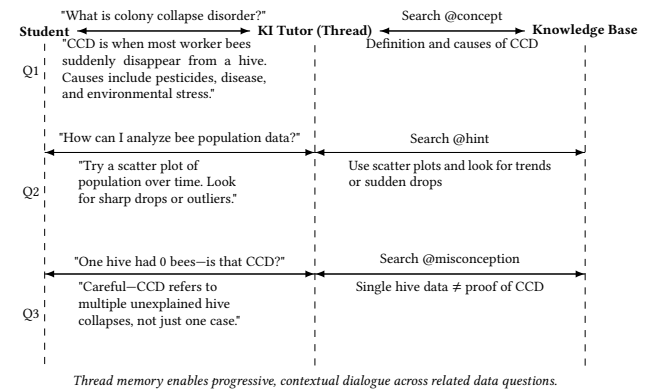
\begin{figure}[h]
\centering
\resizebox{\linewidth}{!}{%
\begin{tikzpicture}[font=\scriptsize, node distance=1.1cm, >=latex]

\node (user) at (0,0) {\textbf{Student}};
\node (tutor) at (4,0) {\textbf{KI Tutor (Thread)}};
\node (kb) at (8,0) {\textbf{Knowledge Base}};

\draw[dashed] (user) -- ++(0,-5);
\draw[dashed] (tutor) -- ++(0,-5);
\draw[dashed] (kb) -- ++(0,-5);

\node at (-0.2,-0.6) {Q1};
\draw[->] (user) -- (tutor) node[midway, above, sloped] {"What is colony collapse disorder?"};
\draw[->] (tutor) -- (kb) node[midway, above, sloped] {Search @concept};
\draw[->] (kb) -- (tutor) node[midway, below, sloped] {Definition and causes of CCD};
\draw[->] (tutor) -- (user) node[midway, below, sloped] {\parbox{3cm}{"CCD is when most worker bees\\suddenly disappear from a hive. Causes include pesticides, disease, and environmental stress."}};

\node at (-0.2,-2.4) {Q2};
\draw[->] (user) ++(0,-1.8) -- ++(4,0) node[midway, above, sloped] {"How can I analyze bee population data?"};
\draw[->] (tutor) ++(0,-1.8) -- ++(4,0) node[midway, above, sloped] {Search @hint};
\draw[->] (kb) ++(0,-1.8) -- ++(-4,0) node[midway, below, sloped] {\parbox{3cm}{Use scatter plots and look for trends\\or sudden drops}};
\draw[->] (tutor) ++(0,-1.8) -- ++(-4,0) node[midway, below, sloped] {\parbox{3cm}{"Try a scatter plot of\\population over time. Look\\for sharp drops or outliers."}};

\node at (-0.2,-4.2) {Q3};
\draw[->] (user) ++(0,-3.6) -- ++(4,0) node[midway, above, sloped] {"One hive had 0 bees—is that CCD?"};
\draw[->] (tutor) ++(0,-3.6) -- ++(4,0) node[midway, above, sloped] {Search @misconception};
\draw[->] (kb) ++(0,-3.6) -- ++(-4,0) node[midway, below, sloped] {Single hive data $\neq$ proof of CCD};
\draw[->] (tutor) ++(0,-3.6) -- ++(-4,0) node[midway, below, sloped] {\parbox{3cm}{"Careful—CCD refers to\\multiple unexplained hive\\collapses, not just one case."}};

\node at (4,-5.2) {\textit{Thread memory enables progressive, contextual dialogue across related data questions.}};

\end{tikzpicture}
}
\caption{Threaded tutor interaction from RockStartIT's bee data science module. The assistant guides from concept to analysis and clarification.}
\label{fig:thread-bee-data}
\end{figure}

Using OpenAI’s thread model, each user session maintains conversational memory across exchanges. This enables the tutor to refer to prior context for coherence and continuity, tailor scaffolding based on learner progress and sustain meaningful multi-turn educational dialogue.

In parallel, the system applies an adaptive content assembly pipeline:
\begin{enumerate}
    \item \textbf{Intent Matching:} Detects learner need (e.g., help, clarification).
    \item \textbf{Tag-Based Retrieval:} Locates pedagogically relevant snippets.
    \item \textbf{Response Construction:} Builds scaffolded responses that are adapted to prior input.
\end{enumerate}

This enables personalized instructional paths and controlled pedagogical responses.  Thread handling via Assistant API enables a "conversational memory" without persistent user tracking.

\subsection{Real-Time Response Streaming}

To enhance engagement and reduce latency, the system uses OpenAI’s streaming API. Rather than waiting for full responses, output is transmitted token-by-token, simulating natural dialogue pacing:
\begin{itemize}
    \item \textbf{Reduces latency:} Feedback appears instantly, improving flow.
    \item \textbf{Simulates live interaction:} Output feels human and reactive.
    \item \textbf{Boosts attention:} Unfolding answers encourage continued focus.
\end{itemize}

This reinforces the tutor’s responsiveness and real-time support behavior. This design choice makes the AI feel more human and responsive compared to static full-message replies.

\subsection{Novelty and Strengths of the Approach}

The KI Tutor introduces a lightweight yet powerful blueprint for integrating LLMs into educational contexts. Its core novelty lies in harmonizing the flexibility of generative AI with structured pedagogical control—achieving both adaptability and instructional safety without requiring model fine-tuning.

\paragraph{Introduced improvements include:}
\begin{itemize}
    \item \textbf{Curriculum-Constrained Output:} Responses have been grounded in a predefined, structured knowledge base, eliminating hallucinations and ensuring factual and curricular alignment.
    \item \textbf{Instructionally Tuned Prompts:} Behavior is explicitly shaped through domain-specific system instructions, guiding the AI to scaffold learning, avoid solution-giving, and engage students in reflection.
    \item \textbf{Context-Aware Dialogue:} Thread-based memory supports coherent, multi-turn conversations that adapt to learner progress and input history.
    \item \textbf{Live Response Streaming:} Token-level response streaming creates a dynamic, conversational experience, enhancing student engagement and reducing perceived latency.
\end{itemize}

\paragraph{Architectural strengths include:}
\begin{itemize}
    \item \textbf{Transparency and Traceability:} A modular, tagged knowledge base allows educators to inspect and adjust content-response mappings, fostering explainability and trust.
    \item \textbf{Scalable Design:} New topics or modules can be integrated by adding tagged content—without modifying traversal logic or retraining the model.
    \item \textbf{Pedagogical Controllability:} Semantic tags and traversal strategies act as interpretable levers to steer the AI toward didactically sound behaviors.
    \item \textbf{Contextual Personalization:} The tutor dynamically assembles responses based on query intent, learner history, and content tags—enabling fine-grained instructional adaptation.
\end{itemize}

Using structured content control, behavior-aware prompting, and contextual thread management, the KI Tutor delivers an interpretable, extensible, and pedagogically robust tutoring experience. The approach is applicable beyond a single course or age group, offering a reusable framework for safe and effective AI integration in education.


\section{Evaluation}
\label{sec:evaluation}

To assess the functionality and usefulness of the AI tutor, we conducted an empirical evaluation with students and teachers, using the tutor as a study companion during a RockStartIT course.
The aim of this study is to evaluate the acceptance and use of the AI tutor by teachers and learners, using the Technology Acceptance Model (TAM) proposed by~\citet{davis_perceived_1989}.
Special attention is given to how the AI tutor is perceived, the factors that influence its acceptance, and the extent to which it supports the learning process.

\subsection{Methodology}
We employed a mixed-methods approach, collecting quantitative feedback through a TAM-inspired questionnaire and gathering qualitative data via four open-ended questions.
Quantitative items,  12 (plus four additional for teachers) Likert-scale items, measured three core constructs:

\begin{itemize}
  \item \textbf{Perceived Usefulness (PU)} – e.g., "The tutor helped me understand the topic."
  \item \textbf{Perceived Ease of Use (PEOU)} – e.g., "It was easy to ask the tutor questions."
  \item \textbf{Intention to Use (IU)} – e.g., "I would like to use a similar tutor in the future."
\end{itemize}

Each item was rated on a 5-point scale from 1 (strongly disagree) to 5 (strongly agree).
Teachers and learners completed the questionnaire after using the AI tutor in class.
The AI tutor was used for 30–45 minutes on participants' own devices.
They were encouraged to explore freely, ask any questions, and follow their curiosity.
Afterward, they completed a structured questionnaire.

The open-ended questions explored challenges, suggestions for improvement, helpful/unhelpful scenarios, and attitudes toward the tutor's future role in education. 

\subsection{Participants}

\begin{table}[t]
  \centering
  \caption{Participant Overview}
  \label{tab:participants}
  \begin{tabular}{cccc}
    \toprule
    \textbf{Participants} & \textbf{Teachers} & \textbf{Learners} & \textbf{Total} \\
    \midrule
    Total                 & 6                 & 7                 & 13             \\
    Male                  & 5                 & 5                 & 10             \\
    Female                & 1                 & 2                 & 3              \\
    Average age           & 32.5              & 23.7              & 27.8           \\
    \bottomrule
  \end{tabular}
\end{table}

A total of 13 participants took part in the study, including six (prospective) teachers with experience in computer science education and seven learners (see \autoref{tab:participants}), most of whom had no prior programming experience.
Teachers were either course instructors or external computer science educators familiar with the curriculum.
Among the teachers, five were male and one female; among learners, five were male and two female.
The average age was 32.5 years for teachers and 23.7 years for learners.

\paragraph{Demographic Questions:}
\begin{enumerate}
  \item Are you currently working as a teacher or in teacher training?
  \item What gender do you identify with?
  \item How old are you?
\end{enumerate}
These questions were used to analyze responses by demographic groups. The role (teacher vs. learner) determined which follow-up questions were shown.

\subsection{Quantitative Results}
Overall, the AI tutor was rated as easy to use and useful. Students especially appreciated the low threshold for asking questions. \autoref{tab:questionnaire-results} shows a summary of the selected questionnaire results.

\begin{table}[ht]
  \centering
  \caption{Selected Questionnaire Results}
  \label{tab:questionnaire-results}
  \begin{tabular}{p{5.5cm}cc}
    \toprule
    \textbf{Item}                                                        & \textbf{Average} & \textbf{SD} \\ \midrule
    PU8: The AI tutor can be a useful addition to my teaching materials. & 4.5              & 0.54        \\
    PEOU5: I find the AI tutor easy to use.                              & 4.6              & 0.50        \\
    \bottomrule
  \end{tabular}

\end{table}

The responses indicated strong acceptance among learners, with most items scoring above 4. Teachers also rated the tutor positively, noting its potential as a support tool rather than a replacement for instruction.

\subsubsection{Perceived Usefulness}
\label{sec:Evaluation:Results:PerceivedUsefulness}

\begin{figure}[h]
  \centering
  \begin{tikzpicture}
    \begin{axis}[
        xbar stacked,
        bar width=1em,
        width=22em,
        height=15em,
        enlarge y limits=0.2,
        enlarge x limits={abs=1},
        xlabel={Number of responses},
        symbolic y coords={PU5,PU4,PU3,PU2,PU1},
        ytick=data,
        xtick={0,2,4,6,8,10,12},
        xmajorgrids=true,
        legend style={
            at={(6em,-5em)},
            anchor=north,
            legend columns=3,
            /tikz/every even column/.append style={column sep=1em}
          },
        legend image post style={draw=black},
        nodes near coords,
        nodes near coords align={center},
        every node near coord/.append style={
            font=\small,
            color=black,
            anchor=center
          },
        tick label style={font=\small},
        yticklabel style={align=left, text width=3em, font=\small},
        yticklabels={
            \textbf{PU5},
            \textbf{PU4},
            \textbf{PU3},
            \textbf{PU2},
            \textbf{PU1},
          }
      ]

      \addplot+[xbar, fill=kit-red60, draw=black] coordinates {
          (0,PU5) (1,PU4) (0,PU3) (1,PU2) (0,PU1)
        };

      \addplot+[xbar, fill=kit-purple50, draw=black] coordinates {
          (4,PU5) (3,PU4) (2,PU3) (0,PU2) (3,PU1)
        };

      \addplot+[xbar, fill=kit-gray30, draw=black] coordinates {
          (2,PU5) (1,PU4) (3,PU3) (1,PU2) (1,PU1)
        };

      \addplot+[xbar, fill=kit-blue40, draw=black] coordinates {
          (1,PU5) (2,PU4) (5,PU3) (6,PU2) (6,PU1)
        };

      \addplot+[xbar, fill=kit-green60, draw=black] coordinates {
          (6,PU5) (6,PU4) (3,PU3) (5,PU2) (3,PU1)
        };

      \legend{
        strongly disagree,
        disagree,
        neutral,
        agree,
        strongly agree
      }

    \end{axis}
  \end{tikzpicture}
  \caption{Perceived Usefulness (PU) ratings from teachers / learners}
  \label{fig:PU_Gesamt}
\end{figure}
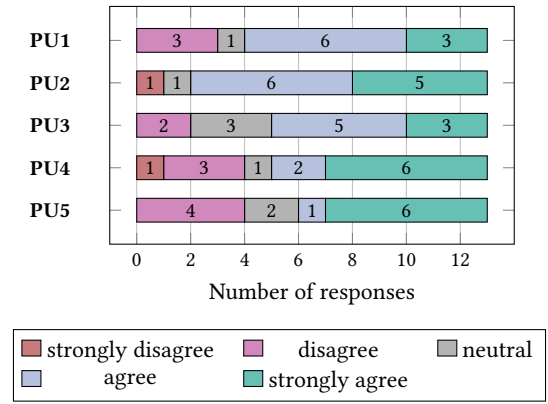

This subsection analyzes participant responses related to \textit{Perceived Usefulness (PU)}. \autoref{fig:PU_Gesamt} shows the distribution of responses across all PU items, and \autoref{tab:pu_stats} summarizes the corresponding descriptive statistics.

Overall, participants rated the AI tutor as useful, particularly in terms of its reliability and potential to support learners. The statement PU3 received the highest level of agreement, followed closely by PU1 and PU2.

Notably, PU1 and PU4 showed a strong positive correlation ($r = 0.77$), while PU2 and PU4 demonstrated a moderate correlation ($r = 0.58$). These results suggest that participants who found the AI tutor reliable and supportive were also more likely to see it as a meaningful aid in completing learning tasks.

Teacher responses were particularly favorable: items PU2 and PU8 received unanimous agreement, highlighting the system's perceived consistency and relevance to classroom practice. However, the statement PU6 received more mixed responses---two out of six teachers expressed slight skepticism. This suggests that while teachers acknowledge the tutor's instructional value, its integration into daily teaching workflows remains an open question.

Learner ratings, though generally positive, displayed greater variance. PU1 and PU2 were the most highly rated, while PU4 and PU5 received more moderate evaluations (mean $M = 3.43$), still trending toward agreement. This variation likely reflects the mismatch between the tutor's intended audience (grades 7–10) and the higher-level academic background of the surveyed university students. Some learners noted that the tutor's responses were too generic or overly cautious---an intentional design choice aimed at promoting reflection rather than simply delivering answers.

In summary, the AI tutor was perceived as useful, particularly in supporting understanding and providing consistent feedback. Teachers in particular valued its reliability and potential to complement instruction, though questions remain about its direct utility in streamlining their workload. Learners appreciated its clarity and guidance, though the design trade-off between answer provision and Socratic scaffolding led to mixed reactions. These findings underline the importance of aligning AI tutor design with both pedagogical goals and the specific needs of target user groups.

\begin{table*}[h!]
  \centering
  \caption{Descriptive statistics for PU, PEOU and IU items}
  \label{tab:pu_stats}
  \begin{tabular}{p{13cm} r r r r}
    \toprule
    \textbf{Item}                                                                                      & \multicolumn{1}{c}{\textbf{N}} & \multicolumn{1}{c}{\textbf{Mean}} & \multicolumn{1}{c}{\textbf{Median}} & \multicolumn{1}{c}{\textbf{SD}} \\
    \midrule
    \textbf{PU1}: The AI tutor helped me understand the topics more easily.                                     & 13         & 3.69          & 4               & 1.11        \\
    \hspace{1em}\textit{Teachers only}                                                                             & 6          & 3.67          & 4               & 1.03        \\
    \hspace{1em}\textit{Learners only}                                                                             & 7          & 3.71          & 4               & 1.25        \\
    \textbf{PU2}: The AI tutor responded reliably to my inputs.                                                 & 13         & 4.08          & 4               & 1.12        \\
    \hspace{1em}\textit{Teachers only}                                                                             & 6          & 4.50          & 4.5             & 0.55        \\
    \hspace{1em}\textit{Learners only}                                                                             & 7          & 3.71          & 4               & 1.38        \\
    \textbf{PU3}: The AI tutor improved the learning experience.                                                & 13         & 3.69          & 4               & 1.03        \\
    \hspace{1em}\textit{Teachers only}                                                                             & 6          & 3.83          & 4               & 0.75        \\
    \hspace{1em}\textit{Learners only}                                                                             & 7          & 3.57          & 4               & 1.27        \\
    \textbf{PU4}: I think the AI tutor can meaningfully support learners with tasks.                            & 13         & 3.69          & 4               & 1.49        \\
    \hspace{1em}\textit{Teachers only}                                                                             & 6          & 4.00          & 4.5             & 1.27        \\
    \hspace{1em}\textit{Learners only}                                                                             & 7          & 3.43          & 4               & 1.71        \\
    \textbf{PU5}: I find the AI tutor useful.                                                                   & 13         & 3.69          & 4               & 1.38        \\
    \hspace{1em}\textit{Teachers only}                                                                             & 6          & 4.00          & 4.5             & 1.27        \\
    \hspace{1em}\textit{Learners only}                                                                             & 7          & 3.43          & 3               & 1.51        \\
    \textbf{PU6}: I believe an AI tutor like this can be useful in my work/teaching. \textit{(teachers only)}   & 6          & 3.50          & 4               & 1.23        \\
    \textbf{PU7}: Using such an AI tutor in class makes my job easier. \textit{(teachers only)}                 & 6          & 3.67          & 3.5             & 1.21        \\
    \textbf{PU8}: The AI tutor can be a useful addition to my teaching materials. \textit{(teachers only)}      & 6          & 4.50          & 4.5             & 0.55        \\
    \midrule
    \textbf{PEOU1}: Interacting with the AI tutor is clear and understandable.                                  & 13         & 4.54          & 5.0             & 0.52        \\
    \hspace{1em}\textit{Teachers only}                                                                             & 6          & 4.12          & 4.0             & 0.41        \\
    \hspace{1em}\textit{Learners only}                                                                             & 7          & 4.86          & 5.0             & 0.38        \\
    \textbf{PEOU2}: I find the AI tutor easy to use overall.                                                    & 13         & 4.62          & 5.0             & 0.51        \\
    \hspace{1em}\textit{Teachers only}                                                                             & 6          & 4.5           & 4.5             & 0.55        \\
    \hspace{1em}\textit{Learners only}                                                                             & 7          & 4.71          & 5               & 0.49        \\
    \textbf{PEOU3}: It is easy to use the AI tutor even without technical knowledge.                            & 13         & 4.46          & 5.0             & 0.66        \\
    \hspace{1em}\textit{Teachers only}                                                                             & 6          & 4.5           & 4.5             & 0.55        \\
    \hspace{1em}\textit{Learners only}                                                                             & 7          & 4.43          & 5               & 0.79        \\
    \textbf{PEOU4}:  The AI tutor does what I want without issues.                                              & 13         & 3.77          & 4.0             & 1.24        \\
    \hspace{1em}\textit{Teachers only}                                                                             & 6          & 3.67          & 4.0             & 1.37        \\
    \hspace{1em}\textit{Learners only}                                                                             & 7          & 3.86          & 4               & 1.22        \\
    \textbf{PEOU5}:  I find the AI tutor easy to use.                                                           & 12         & 4.42          & 4.5             & 0.67        \\
    \hspace{1em}\textit{Teachers only}                                                                             & 5          & 4.6           & 5.0             & 0.55        \\
    \hspace{1em}\textit{Learners only}                                                                             & 7          & 4.29          & 4               & 0.76        \\
    \midrule
    \textbf{IU1}: Assuming I had access to such an AI tutor, I would intend to use it.                          & 13         & 3.85          & 4.0             & 1.41        \\
    \hspace{1em}\textit{Teachers only}                                                                             & 6          & 3.83          & 4.0             & 1.27        \\
    \hspace{1em}\textit{Learners only}                                                                             & 7          & 3.86          & 5               & 1.68        \\
    \textbf{IU2}: If I had access to this AI tutor, I predict I would use it.                                   & 13         & 3.62          & 4.0             & 1.26        \\
    \hspace{1em}\textit{Teachers only}                                                                             & 6          & 3.5           & 3.5             & 1.38        \\
    \hspace{1em}\textit{Learners only}                                                                             & 7          & 3.71          & 4               & 1.25        \\
    \textbf{IU3}: I can imagine using such an AI tutor regularly in my work/classroom. \textit{(teachers only)} & 6          & 3.83          & 4.0             & 1.17        \\
    \bottomrule
  \end{tabular}
\end{table*}

\subsubsection{Perceived Ease of Use}
\label{sec:Evaluation:Results:PerceivedEase}

\begin{figure}[h]
  \centering

  \begin{tikzpicture}
    \begin{axis}[
        xbar stacked,
        bar width=1em,
        width=22em,
        height=15em,
        enlarge y limits=0.2,
        enlarge x limits={abs=1},
        xlabel={Number of responses},
        symbolic y coords={PEOU5,PEOU4,PEOU3,PEOU2,PEOU1},
        ytick=data,
        xtick={0,2,4,6,8,10,12},
        xmajorgrids=true,
        legend style={
            at={(6em,-5em)},
            anchor=north,
            legend columns=3,
            /tikz/every even column/.append style={column sep=1em}
          },
        legend image post style={draw=black},
        nodes near coords,
        nodes near coords align={center},
        every node near coord/.append style={
            font=\small,
            color=black,
            anchor=center
          },
        tick label style={font=\small},
        yticklabel style={align=left, text width=3em, font=\small},
        yticklabels={
            \textbf{PEOU5},
            \textbf{PEOU4},
            \textbf{PEOU3},
            \textbf{PEOU2},
            \textbf{PEOU1},
          }
      ]

      \addplot+[xbar, fill=kit-red60, draw=black] coordinates {
          (0,PEOU5) (0,PEOU4) (0,PEOU3) (0,PEOU2) (0,PEOU1)
        };

      \addplot+[xbar, fill=kit-purple50, draw=black] coordinates {
          (0,PEOU5) (3,PEOU4) (0,PEOU3) (0,PEOU2) (0,PEOU1)
        };

      \addplot+[xbar, fill=kit-gray30, draw=black] coordinates {
          (1,PEOU5) (2,PEOU4) (1,PEOU3) (0,PEOU2) (0,PEOU1)
        };

      \addplot+[xbar, fill=kit-blue40, draw=black] coordinates {
          (5,PEOU5) (3,PEOU4) (5,PEOU3) (5,PEOU2) (6,PEOU1)
        };

      \addplot+[xbar, fill=kit-green60, draw=black] coordinates {
          (6,PEOU5) (5,PEOU4) (7,PEOU3) (8,PEOU2) (7,PEOU1)
        };


    \end{axis}
  \end{tikzpicture}

  \caption{Perceived Ease of Use (PEOU) ratings}
  \label{fig:PEOU_Gesamt}
\end{figure}
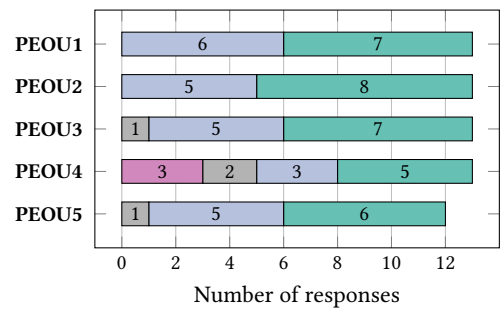

Participants generally rated the AI tutor as easy to use (see \autoref{fig:PEOU_Gesamt}). The most positively rated items were PEOU1 and PEOU2. The only item to receive notable disagreement was PEOU4, which reflects a degree of variability in participants' perceptions of system responsiveness.

Learners tended to rate ease of use slightly more favorably than teachers, though both groups expressed consistent agreement with most statements. Among learners, PEOU1 received the strongest endorsement, with six participants selecting the highest level of agreement. Among teachers, all items except PEOU4 were rated positively, while PEOU4 drew two slightly negative responses, indicating occasional mismatches between user intent and system behavior.

Overall, the AI tutor was perceived as intuitive and accessible. Participants found the chat-based interface familiar and easy to navigate --- especially students, who may be more accustomed to conversational digital tools. Teachers also responded positively, although their slightly lower ratings on PEOU4 suggest that occasional functional limitations may have interfered with expectations of seamless interaction.

These results support the design decision to use a conversational interface. The generally high ratings across both groups indicate that the user interface posed minimal barriers to adoption. The system's clarity and straightforward interaction style contributed to a strong perception of ease of use---an essential factor for engagement, particularly in educational environments where user confidence and comfort directly impact learning outcomes.

\subsubsection{Intention to Use}
\label{sec:Evaluation:Results:IntentionToUse}

\begin{figure}[h]
  \centering
  \begin{tikzpicture}
    \begin{axis}[
        xbar stacked,
        bar width=1em,
        width=22em,
        y=3em,
        enlarge y limits=0.4,
        enlarge x limits={abs=1},
        xlabel={Number of responses},
        symbolic y coords={IU2,IU1},
        ytick=data,
        xtick={0,2,4,6,8,10,12},
        xmajorgrids=true,
        legend style={
            at={(6em,-5em)},
            anchor=north,
            legend columns=3,
            /tikz/every even column/.append style={column sep=1em}
          },
        legend image post style={draw=black},
        nodes near coords,
        nodes near coords align={center},
        every node near coord/.append style={
            font=\small,
            color=black,
            anchor=center
          },
        tick label style={font=\small},
        yticklabel style={align=left, text width=3em, font=\small},
        yticklabels={
            \textbf{IU2},
            \textbf{IU1}
          }
      ]

      \addplot+[xbar, fill=kit-red60, draw=black] coordinates {
          (0,IU2) (1,IU1)
        };

      \addplot+[xbar, fill=kit-purple50, draw=black] coordinates {
          (4,IU2) (2,IU1)
        };

      \addplot+[xbar, fill=kit-gray30, draw=black] coordinates {
          (1,IU2) (1,IU1)
        };

      \addplot+[xbar, fill=kit-blue40, draw=black] coordinates {
          (4,IU2) (3,IU1)
        };

      \addplot+[xbar, fill=kit-green60, draw=black] coordinates {
          (4,IU2) (6,IU1)
        };


    \end{axis}
  \end{tikzpicture}

  \caption{Intention to Use (IU) ratings from teachers / learners}
  \label{fig:IU_Gesamt}
\end{figure}
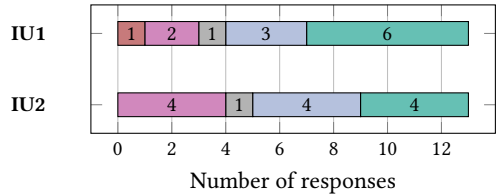

\autoref{fig:IU_Gesamt} summarizes participants' responses to items related to \textit{Intention to Use (IU)}. Both items---IU1 and IU2---were rated positively by the majority of participants, indicating a general openness toward future use. IU1 received slightly stronger overall agreement, though it also drew the only strongly negative response in this category.

Among learners, responses were slightly more variable, reflecting differing levels of enthusiasm. In contrast, teachers provided more consistently positive ratings. The teacher-specific item IU3 also received strong agreement, further reinforcing the system's perceived practical utility.

A notable finding is the strong positive correlation ($r = 0.79$) between participants' perception of system reliability (PEOU4) and their stated intention to use the tutor in the future (IU2). This suggests that perceived consistency in system behavior plays a crucial role in shaping user trust and the likelihood of adoption. Additionally, among teachers, a moderate positive correlation ($r = 0.48$) emerged between age and IU3, indicating that older educators in the sample were slightly more inclined to envision regular classroom use --- possibly reflecting a greater appreciation for tools that support workload management.

Overall, the intention to use the AI tutor was high across both groups. Students showed enthusiasm likely linked to their digital fluency and comfort with conversational interfaces. Teachers' responses were especially encouraging: despite their critical role as gatekeepers of classroom technology, most expressed a clear willingness to integrate the tutor into their instructional practice. Open-ended responses echoed these sentiments, highlighting the tutor's value for delivering one-on-one support in settings with limited teaching staff.

However, several participants also cautioned that the tutor's effectiveness depends heavily on pedagogical alignment. In rigidly structured, answer-driven classrooms, its open-ended guidance may offer limited benefit. This underscores the importance of flexible deployment strategies that empower teachers to tailor AI support to diverse instructional contexts and goals.

\subsubsection{Qualitative Feedback}
\label{sec:Evaluation:Results:Qualitative}

The following open-ended questions were designed to complement the TAM constructs and provide more in-depth insight into participant experiences.

\begin{enumerate}[label=Q\arabic*, left=1em]
  \item \textbf{What challenges did you encounter while using the AI tutor?}
        Targets perceived ease of use (PEOU), exploring usability obstacles and user experiences.

  \item \textbf{Do you have suggestions for how the AI tutor could be improved?}
        Gathers direct feedback for iterative development and adaptation to user needs.

  \item \textbf{Were there situations where the AI tutor was especially helpful or less helpful?}
        Addresses perceived usefulness (PU) and helps identify strong or weak use cases.

  \item \textbf{Do you think AI tutors like this could be a useful addition to future teaching? Why or why not?}
        Aims to understand long-term acceptance and perceived value (IU), especially from the perspective of educators.
\end{enumerate}

\noindent
\textbf{Representative Quotes:}
\begin{itemize}[left=1ex]
  \item \textit{Q\textsubscript{1} "It is sometimes unclear what the tutor's level of knowledge is."}
  \item \textit{Q\textsubscript2: "Example questions that one can ask."}
  \item \textit{Q\textsubscript3: "Was very helpful in summarizing the entire content."}
  \item \textit{Q\textsubscript3: "Especially helpful: clarifying unknown terms through paraphrasing, obtaining additional information."}
  \item \textit{Q\textsubscript3: "It helped me understand why I answered questions incorrectly. That way I could trace back where my thinking error was."}
  \item \textit{Q\textsubscript4: "Yes, absolutely. Using the AI tutor is equivalent to one-on-one support, which is not possible through a teacher in regular classes. It relieves the teacher and enriches the students."}
  \item \textit{Q\textsubscript4: "...already asking the 'right' questions requires a deeper engagement with the content."}
\end{itemize}

The open-ended responses provided rich insights into participants' experiences, expectations, and suggestions for the AI tutor. Feedback revealed a strong interest in improving the system's usability, transparency, and instructional value---alongside an overall positive perception of its educational potential.

\paragraph{Usability and Onboarding.}
One of the most consistent themes was the need for improved onboarding. Both learners and teachers reported uncertainty regarding how to interact effectively with the tutor. Participants suggested the inclusion of example prompts, guided walkthroughs, and clearer indications of the tutor's capabilities and limitations. Teachers, in particular, emphasized the importance of managing expectations by clearly defining what the AI can and cannot do from the outset.

\paragraph{Context Awareness and Curriculum Integration.}
Several responses highlighted the tutor's limited understanding of the lesson context, which occasionally resulted in generic or repetitive answers. Participants emphasized that deeper curriculum integration---such as synchronizing with current chapters or providing task-specific scaffolding---would enhance the system's relevance and effectiveness. Teachers also expressed interest in features that would allow them to guide or constrain the tutor's focus in alignment with instructional goals.

\paragraph{Instructional Support and Modalities.}
Many users appreciated the tutor's ability to explain concepts, offer alternative perspectives, and clarify misunderstandings---especially for novice learners. However, there was a clear call for more example-based explanations and multimodal support, such as the ability to interpret images or present graphical explanations. Participants noted that such features would improve accessibility and support a broader range of learning styles. Teachers also valued the tutor's capacity to support learners with reading or writing difficulties, citing its robust handling of minor input errors.

\paragraph{Autonomy and Engagement.}
Teachers and students alike recognized the AI tutor's potential to foster independent learning and reduce the need for constant teacher supervision---particularly in large or mixed-ability classrooms. The system was perceived as emotionally safe, encouraging learners to ask questions freely without fear of judgment. Several comments noted that this availability and neutrality helped build confidence, especially among learners who were less confident.

\paragraph{Concerns and Cautions.}
Despite the generally positive reception, participants raised valid concerns. Some cautioned that the tutor might encourage surface-level engagement if learners relied too heavily on direct answers rather than developing problem-solving strategies. Others noted occasional repetitiveness or ambiguity in responses. Teachers were especially aware of the risk of over-reliance.
They advocated for a complementary integration model, where the tutor supports but does not replace traditional teaching and group-based learning.

\paragraph{Design Implications.}
Taken together, the feedback points to a range of short-term and long-term development priorities. In the near term, improvements should focus on onboarding, prompt design, and interface clarity. In the long term, participants called for deeper curricular integration, adaptive capabilities, and multimodal interaction (e.g., image input or voice response). These enhancements could significantly strengthen the tutor's educational impact and ensure its alignment with diverse teaching practices.

Overall, qualitative responses highlighted the strengths of a tutor in clarity, availability, inclusiveness, and encouragement of learner autonomy, providing actionable insights for refinement. The findings reinforce the potential of AI tutors to complement human instruction and underscore important design considerations for their effective adoption in educational settings.
\section{Discussion}
\label{sec:Discussion}

At the heart of this project lies a pragmatic philosophy: to design an AI tutor that complements rather than replaces human instruction, built not through costly fine-tuning but through prompt engineering and structured content curation. The result is a system that offers pedagogical guidance, respects curricular boundaries, and can be deployed in real classrooms today—not just in theory.

\subsection{Perceived Value and Role of the Tutor}

Despite a modest sample size (13 participants), the evaluation provided meaningful insight into how such a tutor might integrate into authentic educational settings. Students valued the opportunity to ask questions freely and receive immediate, supportive responses. Teachers appreciated the clarity and consistency of the tutor’s output, especially for reinforcing or rephrasing core concepts. In both groups, the tutor was seen not as a replacement for teaching but as a complementary support—available on demand and reliably focused on the curriculum~\cite{holstein2019studentai, winkler2020chatboted}.

This consistently constrained behavior was among the tutor’s most appreciated features. Unlike generic chatbots, it neither drifted off-topic nor overwhelmed students with irrelevant or overly complex content. Instead, it followed a deliberately modest script: scaffold questions, explain concepts, encourage effort, and refrain from solving tasks outright. This made interactions feel safe, trustworthy, and pedagogically coherent.

\subsection{From TAM to Trust and Pedagogy}

While the Technology Acceptance Model (TAM) provided a solid analytical lens for measuring perceived usefulness and ease of use~\cite{davis_perceived_1989}, feedback revealed important dimensions beyond its scope. One notable concern was \emph{trust calibration}: some students were frustrated when the tutor refused to provide direct answers, while others relied on its output uncritically~\cite{lee2004trust, yin2019understanding}. This dual challenge is well-documented in human-AI interaction and must be addressed through transparency, uncertainty signaling, and strategies that scaffold reflection.

Another theme that emerged was the tutor’s influence on learning behavior. While many participants appreciated how it encouraged inquiry and reflection, some expressed concern that it might foster surface-level engagement or passive question-asking. These findings point to a need for integrating constructs such as \emph{epistemic trust}, \emph{learner agency}, and \emph{pedagogical alignment} into future AI tutor evaluations.

\subsection{Implications for SE and STEM Education}

Although deployed in a general computing course, the tutor’s design principles translate well to software engineering (SE) education, where learners often work independently, face ill-structured problems, and require just-in-time conceptual and technical support~\cite{gross2022aiSEsupport, robins2010learning}.

Key lessons include the value of \emph{instructionally tuned prompting}~\cite{liu2023prompt}, the importance of modular, tagged knowledge bases for precision and scalability, and the tutor’s ability to scaffold learner autonomy through persistent, low-pressure guidance~\cite{lister2004multi}. Its threaded memory and traversal logic enable adaptive support across tasks, while its refusal to provide direct solutions helps maintain the productive struggle essential to deep learning~\cite{chi2008observing}.

The tutor also demonstrates how AI can serve as a \emph{metacognitive partner}, guiding learners through hypothesis testing, debugging, and design reasoning—all skills central to software engineering practice~\cite{azevedo2013handbook}. Importantly, logging student-AI interactions may support formative assessment by identifying misconceptions and guiding instructional interventions~\cite{holmes2019ai}.

\subsection{Reframing AI Tutoring: From Disruption to Enablement}

Our findings support a shift from disruptive hype to grounded enablement. Echoing critiques by Reich and others~\cite{reich2020failure}, this tutor does not seek to transform education overnight. Instead, it extends access to individualized support in contexts where human help is limited. As such, it amplifies rather than replaces existing pedagogical practices.

Its novelty lies not in technical complexity but in deliberate constraint. It is safe, interpretable, curriculum-aligned, and easy to extend—qualities often overlooked in AI deployments. This makes it a viable template for educational use in schools, even those without deep technical infrastructure.

Rather than replacing classroom instruction, the AI tutor supports it—acting as a bridge between human mentorship and student independence. In doing so, it offers a promising pathway toward more inclusive, adaptive, and pedagogically grounded computing education.

\section{Limitations}
\label{sec:Limitations}

While this exploratory study offers promising evidence for the feasibility and acceptance of AI tutors in computing education, several limitations constrain the scope and generalizability of our findings.

First, the small sample size (13 participants) and the mismatch between the test group (primarily male university students and pre-service teachers) and the intended target group (secondary school learners and teachers) limit external validity. Participants’ higher digital literacy and maturity may have positively biased their experience, particularly regarding usability and interaction style~\cite{xu2022aiinclassroom}.

Second, the AI tutor was evaluated as an early-stage prototype in a standalone web environment. Its integration into the full RockStartIT platform was not yet realized, which affected contextual alignment and task continuity. Moreover, the evaluation sessions were short and one-off; without longitudinal data, we cannot assess how user trust, engagement, or learning strategies evolve over time~\cite{roschelle2000advancing}.

Third, the system’s current capabilities were constrained by platform limitations—such as lack of persistent memory, multimodal input, or fine-grained user modeling—hindering deeper personalization and task-specific support. In addition, no behavioral usage data (e.g., interaction logs, time-on-task) were collected, making it difficult to triangulate self-reports with actual learning behavior.

Finally, while most participants rated the tutor positively, the study did not explicitly assess trust calibration. As previous research shows, both over-reliance and under-reliance on AI pose educational risks~\cite{bakhtiar2022trusteducationalai}. Future evaluations must examine how learners engage critically with AI feedback and how trust can be scaffolded through interaction design.

\section{Conclusion}
\label{sec:Conclusion}

This paper presented the design and pilot evaluation of a GPT-based AI tutor developed to support secondary-level learners in a digital software engineering course. By embedding the tutor into a real-world learning context and evaluating it with both students and teachers, we observed high levels of perceived usefulness, usability, and intention to use.

Our findings show that conversational AI, when carefully constrained and pedagogically structured, can enhance learner autonomy, engagement, and reflective thinking. The tutor’s modular knowledge base—paired with instructionally aligned prompts --- enabled scaffolded support while avoiding direct solution delivery, thus promoting inquiry rather than answer-seeking.

These results underscore the potential of AI tutors to complement human instruction, particularly in settings where real-time feedback is limited. The positive reception from both students and educators suggests a viable role for such systems in future software engineering education.

Future work should focus on deployment with the intended target group in authentic classroom settings, where language needs and attention spans differ significantly from adult learners. Longitudinal studies are needed to assess sustained engagement, trust calibration, and learning gains. Technical development will involve tighter integration with the RockStartIT platform, synchronization with learner progress, and support for multimodal inputs (e.g., code snippets, diagrams).

Pedagogically, future versions should explore adaptive scaffolding techniques such as progressive hinting, reflective prompts, and feedback on misconceptions. Designing for trust—through strategies like uncertainty signaling or response justification—can further encourage critical thinking and epistemic agency. Improved onboarding and voice-enabled interaction may increase accessibility for younger learners or students with literacy challenges.

Ultimately, AI tutors like the one presented here offer a lightweight, scalable, and pedagogically grounded approach to augmenting software engineering education. Realizing this potential will require continued iteration, empirical validation, and close collaboration with educators—ensuring that AI becomes not a replacement, but a trusted partner in teaching and learning.


\bibliographystyle{ACM-Reference-Format}
\bibliography{references, thesis, referencesICSEAI, rw}


\end{document}